\documentclass[a4paper,12pt]{article}
\usepackage{a4wide,latexsym,graphicx,epsfig,psfrag,float}
\usepackage{mathrsfs}
\usepackage{latexsym}
\usepackage{indentfirst}
\usepackage{amsmath}
\usepackage{amsxtra}
\usepackage{amsbsy}
\usepackage{amscd}
\usepackage{color}
\usepackage{bbm}
\usepackage{hyperref}
\usepackage{multirow}

\usepackage{graphics}
\usepackage{amssymb}
\usepackage{amsfonts}

\usepackage{subfigure}
\usepackage{color}
\usepackage{times,fancyhdr}
\usepackage{amsfonts}

\allowdisplaybreaks

\newcommand{\rbox}{\rule[-0.20cm]{0cm}{8mm}}
\newcommand{\be}{\begin{equation}}
\newcommand{\ee}{\end{equation}}

\def\no{\nonumber}
\def\bea{\arraycolsep .1em \begin{eqnarray}}
\def\eea{\end{eqnarray}}

\begin{document}

\begin{titlepage}
%\vskip 1cm
%\begin{flushright}
%{
%JLAB-THY-16-
%}
%\end{flushright}
\vspace{2cm}

\setcounter{footnote}{0}
\renewcommand{\thefootnote}{\fnsymbol{footnote}}

\begin{center}
{\LARGE \bf Pion Polarizabilities from $\gamma\gamma\to\pi\pi$ Analysis }
\vspace{2cm} \\
{\sc  Ling-Yun Dai}$^{1,2,3}$\footnote{Email:~l.dai@fz-juelich.de}
{\sc , M.R. Pennington}$^{4,5}$\footnote{Email:~michaelp@jlab.org}
\vspace{0.8cm} \\
$^{1}$ {\small Institute for Advanced Simulation, Institut f\"ur Kernphysik
   and J\"ulich Center for Hadron Physics, Forschungszentrum J\"ulich, D-52425 J\"ulich,
 Germany} \\[1mm]
$^{2}$ {\small Center For Exploration of Energy and Matter, Indiana University, Bloomington, IN 47408, USA} \\[1mm]
$^{3}$ {\small Physics Department, Indiana University, Bloomington, IN 47405, USA}\\[1mm]
$^{4}$ {\small Thomas Jefferson National Accelerator Facility, Newport News, VA 23606, USA} \\[1mm]
$^{5}$ {\small Physics Department, College of William \& Mary, Williamsburg, VA 23187, USA}\\[1mm]
\end{center}

\setcounter{footnote}{0}
\renewcommand{\thefootnote}{\arabic{footnote}}
%\vspace{0.5cm}

\begin{abstract}
We present results for pion polarizabilities predicted using dispersion relations from our earlier Amplitude Analysis of world data on two photon production of meson pairs.
The helicity-zero polarizabilities are rather stable and insensitive to uncertainties in cross-channel exchanges. The need is first to confirm the recent result on $(\alpha_1-\beta_1)$ for the charged pion by COMPASS at CERN to an accuracy of 10\%  by measuring the $\gamma\gamma\to\pi^+\pi^-$ cross-section to an uncertainty of ~1\%. Then the same polarizability, but for the $\pi^0$, is fixed to be
$(\alpha_1-\beta_1)_{\pi^0}=(-0.9\pm0.2)\times 10^{-4}$ fm$^{3}$.
By analyzing the correlation between uncertainties in the meson polarizability and those in $\gamma\gamma$ cross-sections, we suggest experiments need to measure these cross-sections between $\sqrt{s}\simeq 350$ and 600~MeV. The $\pi^0\pi^0$ cross-section then makes the  $(\alpha_2-\beta_2)_{\pi^0}$ the easiest helicity-two polarizability to determine.
\end{abstract}
\vspace{1.5cm}

{\indent PACS~: 13.40-f, 11.55.Fv, 11.80.Et, 12.39.Fe, 13.60Le \\
\indent Keywords~: Pion Polarizability, Dispersion relations, Partial-wave analysis, \\
\indent \;\;\;\;\;\;\;\;\;\;\;\;\;\;\;\;\;\,Chiral Lagrangian, meson production}

\end{titlepage}

\section{Introduction}\label{sec:Introduction}
There has  long been interest in studying pion electromagnetic polarizabilities~\cite{Holstein1990, MRPJP}:
the electric polarizability $\alpha$ and the magnetic polarizability $\beta$. These characterize the pion's rigidity against deformation in an external electromagnetic field. The pion polarizability may also play an important role~\cite{Engel2012} in the hadronic light-by-light scattering contribution to $(g-2)_\mu$~\cite{Fermilab-g2}.
Compton scattering is the ideal  way to test polarizabilities as the strong interaction is strong and so compacts
quarks and gluons together to form a stiff hadron. Over the years this has motivated both experimental and theoretical effort.
On the theory side, Chiral Perturbation Theory ($\chi$PT) gives predictions calculated first to $\mathcal{O}(p^4)$~\cite{Holstein1990,Donoghue1989,Babusci1992} and  up to $\mathcal{O}(p^6)$ from \cite{Buergi1996,Gasser06}.
On the experimental side, measurements have been made from the pion radiative scattering $\pi^-Z(A)\to\gamma\pi^-Z(A)$ by IHEP in Serpukhov~\cite{Antipov8385},  from radiative photoproduction on hydrogen $\gamma p\to\gamma\pi^+n$ by the  Lebedev Physical Institute~\cite{Aibergenov1986} and MAMI~\cite{Ahrens2005}, and from $\pi^-Ni\to\gamma\pi^-Ni$ with COMPASS~\cite{COMPASS2014prl}.

Recently  a proposal has been accepted to study polarizabilities by measuring low energy $\gamma\gamma\to\pi^+\pi^-$~\cite{Jlabproposal} in Hall~D at Jefferson Lab. The issue is then how well do such measurements determine the pion polarizability: reliability and accuracy. This is the issue we address here. In our previous work~\cite{DLY-MRP14} we made a precise Amplitude Analysis of extant data on $\gamma\gamma\to\pi\pi$, $\overline{K}K$ up to $\sqrt{s}=1.5$GeV, and built a dispersive way to calculate amplitudes in the low energy region. This makes a prediction of pion polarizability possible.
The paper is organized as follows: In Sect.~2 we give the formalism for the underlying amplitudes and their relation to pion polarizabilities.
In Sect.~3 we give our prediction for pion polarizabilities, and consider the correlation between the cross-section and pion polarizability to assess the energy domain where sensitivity is greatest.
Finally we summarize.

\section{Formalism for Pion Polarizabilities}\label{sec:2}
\subsection{Amplitudes}\label{sec:2;1}

\noindent As is well known, pion polarizabilities are determined by how the amplitudes for the Compton scattering, $\gamma\pi\to\gamma\pi$, approach threshold. With Compton scattering in the $t$ and $u$ channels, threshold is the kinematic point $s=0, t=u=m_\pi^2$. While exactly at this threshold the amplitudes are fixed by Low's low energy theorem and given by One Pion Exchange, the deviation from this Born amplitude as $s\to 0$ reflects the rigidity of the pion that are the polarizabilties. By crossing these are, of course, the $\gamma\gamma\to\pi\pi$ amplitudes continued to $s\to 0$~\cite{MRPJP,Gasser06,Filkov05,Filkov06,Pasquini08}. Dispersion relations provide the natural and effective way to continue the $\gamma\gamma$ amplitude analytically to this unphysical region. Here we use the partial wave dispersion relation established in \cite{DLY-MRP14}, for ${\mathcal F}^I_{J\lambda}(s)$, the $\gamma\gamma\to\pi\pi$ amplitudes with definite $\pi\pi$ isospin $I$, spin $J$ and two photon helicity $\lambda$.  ${\mathcal B}^I_{J\lambda}(s)$ denote the corresponding Born contributions. Each of the amplitudes ${\mathcal F}(s)$ has a phase $\varphi(s)$.  From these we can define an  Omn$\grave{e}$s function \cite{Omnes1958}
\be\label{eq:Omnes}
\vspace{3mm}
\Omega^I_{J\lambda}(s)=\exp\left(\frac{s}{\pi} \int^\infty_{s_{th}} ds' \frac{\varphi^I_{J\lambda}(s')}{s'(s'-s)}\right) \,\, .
\ee
Then using constraints such as Low's low energy theorem and  the required threshold behaviour, we can write dispersion relations for the partial waves.
These have contributions from the right hand (unitarity) cut (RHC) and from the left hand cut (LHC). The latter is controlled by $t$ and $u$-channel exchanges, both single and multi-particle.
This contribution is determined by the explicit One Pion Exchange Born amplitude, plus the rest which defines a contribution to ${\cal F}^I_{J\lambda}(s)$ we call ${\cal L}^I_{J\lambda}(s)$.

For $S$-wave amplitudes, these have one subtraction usefully taken at $s=0$ by considering $({\mathcal F}(s)\,-\,{\mathcal B}(s))\Omega^{-1}(s)/s$:
\bea\label{eq:F;ampS}
\mathcal{F}^{I}_{00}(s)\;=\;{\mathcal B}^I_S(s)+b^{I} s~\Omega^{I}_{0}(s)
 &+&\frac{s^2~\Omega^{I}_{0}(s)}{\pi}\int_L ds'\frac{{\rm Im}\left[ \mathcal{L}^{I}_{00}(s')\right]\Omega^{I}_{0}(s')^{-1} }{s'^2(s'-s)} \no \\[2.5mm]
              &-&\frac{s^2\;\Omega^{I}_{0}(s)}{\pi}\int_R ds'\frac{{\mathcal B}^I_S(s')\;{\rm Im}\left[ \Omega^{I}_{0}(s')^{-1}\right] }{s'^2(s'-s)}\, .
\eea
where the $\,b^I\,$ (with $I=0, 2$) are subtraction constants given by:
\bea\label{eq:F;bI0I2}
b^{I=0}&=&\frac{\sqrt{3}\Sigma(s_n)-\sqrt{6}\frac{m_\pi}{4\,\alpha\,}(\alpha_1-\beta_1)_{\pi^+}\Omega^{2}_0(s_n)}{\Omega^{0}_0(s_n)+2\Omega^{2}_0(s_n)} \,\, , \no\\
b^{I=2}&=&\frac{-\sqrt{6}\Sigma(s_n)-\sqrt{3}\frac{m_\pi}{4\,\alpha\,}(\alpha_1-\beta_1)_{\pi^+}\Omega^{0}_0(s_n)}{\Omega^{0}_0(s_n)+2\Omega^{2}_0(s_n)}  \,\, , \no\\
\eea
with
\bea\label{eq:F;Sigma}
\Sigma(s)&=&-\sqrt{\frac{1}{3}}\frac{s_n~\Omega^{I=0}(s_n)}{\pi}\left(
                                 \int_R ds'\frac{\sqrt{\frac{2}{3}} B_S(s)  {\rm Im}\left[\Omega^{0}_0(s')^{-1} \right] }{s'^2(s'-s)}
                                +\int_L ds'\frac{{\rm Im}\left[\mathcal{L}^{0}_{00}(s')\right] \Omega^{0}_0(s')^{-1} }{s'^2(s'-s)}
                                  \right)   \nonumber\\
&&+   \sqrt{\frac{2}{3}}\frac{s_n~\Omega^{I=2}(s_n)}{\pi}\left(
                                 \int_R ds'\frac{\sqrt{\frac{1}{3}} B_S(s)  {\rm Im}\left[ \Omega^{2}_0(s')^{-1}\right] }{s'^2(s'-s)}
                                +\int_L ds'\frac{{\rm Im}\left[\mathcal{L}^{0}_{00}(s')\right] \Omega^{2}_0(s')^{-1} }{s'^2(s'-s)}
                                  \right) \,\, .\nonumber
\eea
$s=s_n$ is the position of the Adler zero in the $\gamma\gamma\to\pi^0\pi^0$ $S$-wave. It's position is at $s_n=(1\pm 0.2)m_{\pi^0}^2$, from ChPT.
For waves with higher spin, {\it i.e} $J\, >\, 0$,
we write
unsubtracted dispersion relations for $({\mathcal F}(s)\,-\,{\mathcal B}(s))\Omega^{-1}(s)/s^n(s\,-\,4m_\pi^2)^{J/2}$:
\bea\label{eq:F;ampJ}
\mathcal{F}^I_{J\lambda}(s)\;=\;{\mathcal B}^I_{J\lambda}(s)
&+&\frac{s^n(s-4m_\pi^2)^{J/2}}{\pi}\,\Omega^{I}_{J}(s)\,\int_L ds'\frac{{\rm Im}\left[ \mathcal{L}^{I}_{J\lambda}(s')\right]\, \Omega^{I}_{J}(s')^{-1} }{s'^n(s'-4m_\pi^2)^{J/2}(s'-s)}\nonumber\\[3.5mm]
&-&\frac{s^n(s-4m_\pi^2)^{J/2}}{\pi}\,\Omega^{I}_{J}(s)\,\int_R ds'\frac{B^{I}_{J\lambda}(s')\,  {\rm Im}\left[ \Omega^{I}_{J}(s')^{-1}\right] }{s'^n(s'-4m_\pi^2)^{J/2}(s'-s)} \,\, ,
\eea
where $n\,=\,2\,-\,\lambda/2$.
As we will discuss later, the polarizabilities are related to $b^I$ and $\mathcal{R}^{I}_{J\lambda}(s)$ (see Eq.~(\ref{eq:Po;R},\ref{eq:RL;fit})).

\subsection{Left Hand Cut Contribution from Single Particle Exchange}\label{sec:2;2}
\begin{figure}[htbp]
\vspace*{-1.5cm}
\includegraphics[width=0.95\textwidth,height=0.95\textheight]{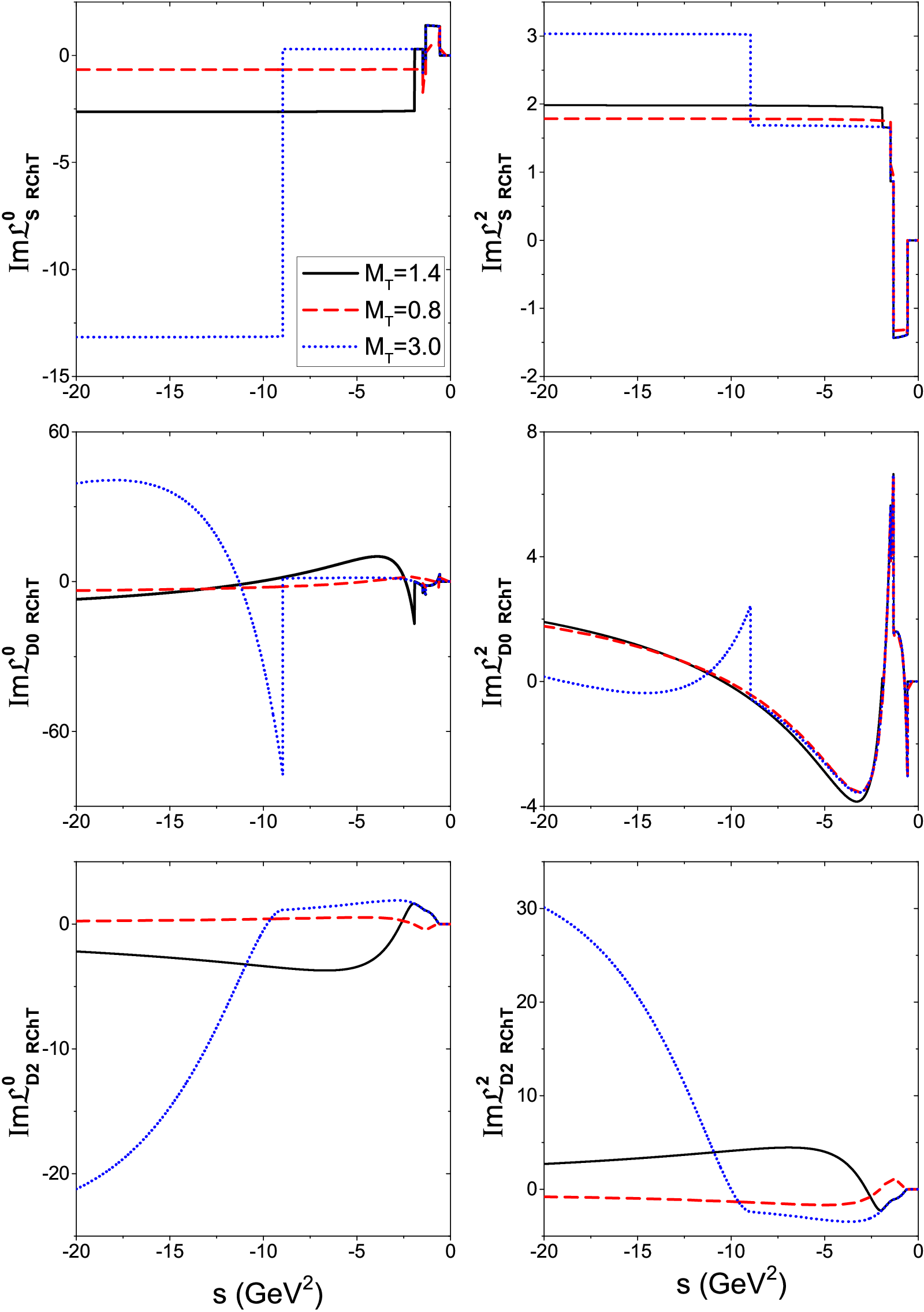}
\caption{\label{fig:ImL;RChT} Left Hand Cut modelled by single particle exchanges $\rho$, $\omega$, $b_1$, $h_1$, $a_1$ and the tensor $T$. The mass of the \lq effective' tensor resonance ($M_T$) is set to 1.4, 0.8, 3.0 GeV for the solid black, dashed red, and dotted blue lines, respectively.      }
\end{figure}

\noindent An idea of what the Left Hand Cut looks like can be estimated by considering single particle exchanges~\cite{DLY-MRP14,Oller0708,yumao09,Moussallam10}.
Of course, single particle exchange in the $\gamma\gamma$ process is a resonance in Compton scattering. We list the imaginary parts, required in evaluating
Eqs.~(\ref{eq:F;ampS},\ref{eq:F;ampJ}), from $\rho$, $\omega$, $b_1$, $h_1$, $a_1$   and an effective tensor resonance $T$:
\bea\label{eq:L;RChT}
{\rm Im}\mathcal{L}^{0~R\chi T}_{J\lambda}(s) &=&
-\sqrt{\frac{3}{2}}{\rm Im}\mathcal{L}_{\rho,J\lambda}(s)-\sqrt{\frac{1}{6}}{\rm Im}\mathcal{L}_{\omega,J\lambda}(s)
-\sqrt{\frac{3}{2}}{\rm Im}\mathcal{L}_{b_1,J\lambda}(s) \no \\
&&-\sqrt{\frac{1}{6}}{\rm Im}\mathcal{L}_{h_1,J\lambda}(s)-\sqrt{\frac{2}{3}}{\rm Im}\mathcal{L}_{a_1}(s)
+{\rm Im}\mathcal{L}_{T,J\lambda}(s)\;,\no \\
{\rm Im}\mathcal{L}^{2~R\chi T}_{J\lambda}(s)&=&\sqrt{\frac{1}{3}}{\rm Im}\mathcal{L}_{\omega,J\lambda}(s)
+\sqrt{\frac{1}{3}}{\rm Im}\mathcal{L}_{h_1,J\lambda}(s)
-\sqrt{\frac{1}{3}}{\rm Im}\mathcal{L}_{a_1,J\lambda}(s)+{\rm Im}\mathcal{L}_{T,J\lambda}(s) \,\, ,
\eea
where, with $M_R$, the mass of the resonance in the Compton channel,
\bea \label{eq:ImL;RChT}
{\rm Im}\mathcal{L}_{R,S}(s)&=&-{ N^R_{J\lambda}\, \pi C_R^{\,2} M_R^{\,2}}/{\rho(s)}\;,\no \\[3mm]
{\rm Im}\mathcal{L}_{R,D0}(s)&=&{\sqrt{5}N^R_{J\lambda}\, \pi C_R^{\,2} M_R^{\,2}}\,[1-3X^2(M_R,s)]/{2\rho(s)}\;,\no \\[3mm]
{\rm Im}\mathcal{L}_{R,D2}(s)&=&{\sqrt{30}N^R_{J\lambda}\, \pi C_R^{\,2} s \rho(s)}\,[1-X^2(M_R,s)]^2/16\;,
\eea
and
\bea\label{eq:X;rho}
X(M,s)&=&\frac{2M^2-2m_\pi^2+s}{s~\rho(s)}\;,\; {\mathrm{with}}\quad \rho(s)=\sqrt{1-4m_\pi^2/s}\; .
\eea
Note that the normalization factors $N^R_{J\lambda}$ are as follows:
\bea
&&N^\omega_{J0,J2}=1\;,\;\; N^\rho_{J0,J2}=\frac{1}{9}\;,\;\;
  N^{a_1}_{J 0}=\frac{1}{4}\;,\;\; N^{a_1}_{J2}=-\frac{1}{4}\;, \no \\
&&N^{b_1}_{J 0}=-\frac{1}{36}\;,\;\; N^{b_1}_{J2}=\frac{1}{36}\;,\;\;
  N^{h_1}_{J 0}=-\frac{1}{4}\;,\;\; N^{h_1}_{J2}=\frac{1}{4}\;. \no
\eea
 The coefficients of $C_R$ are fixed from the
decay widths $R\to\pi\gamma$~\cite{DLY-MRP14}. The couplings of  the effective $T$-exchange are fixed by demanding the sum of the exchange contributions
 cancel when $s\to\infty$. This is why $C_T^{\,2}$ can be negative.
\bea
&C_\rho=1.25\pm0.08,\;\;C_\omega=1.15\pm0.02,\;\;C_{a_1}=1.08\pm0.21,\;\;C_{b_1,h_1}=1.95\pm0.25,\;\; \no\\[2mm]
&C^{\,2}_T(0S0)=0.477,\;\;C^{\,2}_T(0D0)=1.403,\;\;C^{\,2}_T(0D2)=0.354,\;\;\no\\[2mm]
&C^{\,2}_T(2S0)=-0.048,\;\;C^{\,2}_T(2D0)=-0.053,\;\;C^{\,2}_T(2D2)=-0.509,\;\;\no
\eea
with $C_R$ in units of GeV$^{-1}$. The resulting left hand cut terms are then shown in Fig.~\ref{fig:ImL;RChT}.
Changing the mass of the effective resonance $T$ from 0.8 to 3.0~GeV, the left hand cut contributions vary little for the isospin two $S$-waves and $D_0$ waves.
This is a consequence of the coefficients $C_T^{\,2}$ being rather small for these two waves.
The difference in contributions is shown in Fig.~\ref{fig:ImL;RChT}.

\subsection{Pion Polarizabilities}\label{sec:2;3}
\noindent From our two photon partial wave amplitudes, we have scattering amplitudes for $\gamma\gamma\to\pi\pi$
\bea
M_{++}(s, \theta, \phi)&=&e^2\, \sqrt{16\pi}\,  \sum_{J\geq0}\,F_{J0}(s)\,Y_{J0}(\theta,\phi) \,\, , \nonumber\\
M_{+-}(s, \theta, \phi)&=&e^2\, \sqrt{16\pi}\, \sum_{J\geq2}\,F_{J2}(s)\,Y_{J2}(\theta,\phi) \,\, , \label{eq:M}
\eea
with
\bea
Y_{Jm}(\theta,\phi)=\sqrt{\frac{(2J+1)(J-|m|)!}{4\pi(J+|m|)!}}\,P^{|m|}_J(\cos \theta)\,e^{im\phi}
\eea
where $\theta$ and $\phi$ are the scattering (and azimuthal) angles in the $x-z$ plane. From these amplitudes we form the isospin combinations that correspond to whether the pions are neutral or charged to give $M^{n,c}_{J\lambda}$ respectively.
Continuing these to the unphysical region using the Lorentz invariants $s,t$ relates these at $s=0$ to the polarizabilities,
so that
\bea
M^n_{++}(s,\theta=\pi/2,\phi=0)&=&
e^2\sqrt{16\pi}\frac{m_\pi}{4\alpha}\left(s\left(\alpha_1-\beta_1\right)_{\pi^0}+\frac{s^2}{12}\left(\alpha_2-\beta_2\right)_{\pi^0}\right)\,\, , \nonumber\\[1mm]
M^n_{+-}(s,\theta=\pi/2,\phi=0)&=&
e^2\sqrt{16\pi}\frac{m_\pi}{4\alpha}\left(-s\left(\alpha_1+\beta_1\right)_{\pi^0}-\frac{s^2}{12}\left(\alpha_2+\beta_2\right)_{\pi^0}\right)\,\, , \nonumber\\[1mm]
M^c_{++}(s,\theta=\pi/2,\phi=0)&=&
e^2\sqrt{16\pi}\left(B_{++}+\frac{m_\pi}{4\alpha}[s\left(\alpha_1-\beta_1\right)_{\pi^+}+\frac{s^2}{12}\left(\alpha_2-\beta_2\right)_{\pi^+}]\right)\,\, , \nonumber\\[1mm]
M^c_{+-}(s,\theta=\pi/2,\phi=0)&=&
e^2\sqrt{16\pi}\left(B_{+-}-\frac{m_\pi}{4\alpha}[s\left(\alpha_1+\beta_1\right)_{\pi^+}+\frac{s^2}{12}\left(\alpha_2+\beta_2\right)_{\pi^+}]\right)\,\, , \label{eq:P;DP}
\eea
Using the dispersive contributions specified by the cross-channel exchanges from Eq.~(\ref{eq:L;RChT}) to define
reduced amplitudes ${\mathcal R}^I_{J\lambda}(s)$ defined in the Appendix, Eqs.~(\ref{eq:RL;fit},\ref{eq:RLB;fit}), we can rewrite
our amplitudes of Eqs.~(\ref{eq:F;ampS},\ref{eq:F;ampJ}) to obtain the polarizabilities. This has already been discussed in \cite{Moussallam10} considering twice or once subtracted dispersion relations, and in \cite{Phillips11} by solving the  Roy-Steiner equations. However, here we only use once subtracted dispersion relations for $S$-waves and unsubtracted ones for $D$-waves. As we will discuss later, this makes it possible to predict the polarizabilities with less unknown constants, and provides a tighter connection between these and the two photon cross-sections. One has\footnote{We note that in the paper \cite{Moussallam10}, they missed the $d^{(I)}$ term of $(\alpha_2+\beta_2)_{\pi^+,\pi^0}^{I}$ in their Eq.~(69), which corresponds to the first two terms in our representation.}:
\bea\label{eq:Po;R}
(\alpha_1-\beta_1)_{\pi^+}&=&\frac{4\alpha}{m_\pi}\left(-\sqrt{\frac{2}{3}}b^{I=0}-\sqrt{\frac{1}{3}}b^{I=2}\right)\,\, , \nonumber\\[5mm]
(\alpha_2-\beta_2)_{\pi^+}&=&\frac{48\alpha}{m_\pi}\left(-\sqrt{\frac{2}{3}}b^{I=0}\frac{d \Omega^{I=0}_{0}(0)}{ds}-\sqrt{\frac{1}{3}}b^{I=2}\frac{d \Omega^{I=2}_{0}(0)}{ds}-\sqrt{\frac{2}{3}}\mathcal{R}^{I=0}_{00}(s)-\sqrt{\frac{1}{3}}\mathcal{R}^{I=2}_{00}(s)\right)\,\, , \nonumber\\[1.2mm]
&+&96\sqrt{5}\alpha m_\pi\left(-\sqrt{\frac{2}{3}}\mathcal{R}^{I=0}_{20}(s)-\sqrt{\frac{1}{3}}\mathcal{R}^{I=2}_{20}(s)\right)\,\, , \nonumber\\[5mm]
(\alpha_1-\beta_1)_{\pi^0}&=&\frac{4\alpha}{m_\pi}\left(-\sqrt{\frac{1}{3}}b^{I=0}+\sqrt{\frac{2}{3}}b^{I=2}\right)\,\, , \nonumber\\[5mm]
(\alpha_2-\beta_2)_{\pi^0}&=&\frac{48\alpha}{m_\pi}\left(-\sqrt{\frac{1}{3}}b^{I=0}\frac{d \Omega^{I=0}_{00}(0)}{ds}+\sqrt{\frac{2}{3}}b^{I=2}\frac{d \Omega^{I=2}_{00}(0)}{ds}-\sqrt{\frac{1}{3}}\mathcal{R}^{I=0}_{00}(s)+\sqrt{\frac{2}{3}}\mathcal{R}^{I=2}_{00}(s)\right)\,\, , \nonumber\\[1.2mm]
&+&96\sqrt{5}\alpha m_\pi\left(-\sqrt{\frac{1}{3}}\mathcal{R}^{I=0}_{20}(s)+\sqrt{\frac{2}{3}}\mathcal{R}^{I=2}_{20}(s)\right)\,\, , \nonumber\\[5mm]
(\alpha_1+\beta_1)_{\pi^+}&=&4\sqrt{30}\alpha m_\pi\left(-\sqrt{\frac{2}{3}}\mathcal{R}^{I=0}_{22}(0)-\sqrt{\frac{1}{3}}\mathcal{R}^{I=2}_{22}(0)\right)\,\, , \nonumber\\[5mm]
(\alpha_2+\beta_2)_{\pi^+}&=&\frac{-12\sqrt{30}\alpha}{m_\pi}\left(-\sqrt{\frac{2}{3}}\mathcal{R}^{I=0}_{22}(0)-\sqrt{\frac{1}{3}}\mathcal{R}^{I=2}_{22}(0)
+4m_\pi^2\sqrt{\frac{2}{3}}\mathcal{R}^{I=0}_{22}(0)\frac{d \Omega^{I=0}_{22}(0)}{ds}\right.\,\no\\[1.2mm]
&+&\left.4m_\pi^2\sqrt{\frac{1}{3}}\mathcal{R}^{I=2}_{22}(0)\frac{d \Omega^{I=2}_{22}(0)}{ds}+4m_\pi^2\sqrt{\frac{2}{3}}\mathcal{R'}^{I=0}_{22}(0)+4m_\pi^2\sqrt{\frac{1}{3}}\mathcal{R'}^{I=2}_{22}(0)\right)\,\, , \nonumber\\[5mm]
(\alpha_1+\beta_1)_{\pi^0}&=&4\sqrt{30}\alpha m_\pi\left(-\sqrt{\frac{1}{3}}\mathcal{R}^{I=0}_{22}(0)+\sqrt{\frac{2}{3}}\mathcal{R}^{I=2}_{22}(0)\right)\,\, , \nonumber\\[5mm]
(\alpha_2+\beta_2)_{\pi^0}&=&\frac{-12\sqrt{30}\alpha}{m_\pi}\left(-\sqrt{\frac{1}{3}}\mathcal{R}^{I=0}_{22}(0)+\sqrt{\frac{2}{3}}\mathcal{R}^{I=2}_{22}(0)
+4m_\pi^2\sqrt{\frac{1}{3}}\mathcal{R}^{I=0}_{22}(0)\frac{d \Omega^{I=0}_{22}(0)}{ds}\right.\,\no\\[1.2mm]
&-&\left.4m_\pi^2\sqrt{\frac{2}{3}}\mathcal{R}^{I=2}_{22}(0)\frac{d\Omega^{I=2}_{22}(0)}{ds}
+4m_\pi^2\sqrt{\frac{1}{3}}\mathcal{R'}^{I=0}_{22}(0)-4m_\pi^2\sqrt{\frac{2}{3}}\mathcal{R'}^{I=2}_{22}(0)\right)\,\, ,
\eea
Notice that for higher partial waves with $J\geq4$, the Born terms are expected to be an adequate approximation  and so they make no contribution to the pion polarizabilities.
While polarizabilities encode the approach to the One Pion Exchange Born amplitude for Compton scattering at threshold, this does not mean it is independent of the Born amplitude. This is because in some key channels it is the modifications to the Born amplitude from the $\pi\pi$ final state interaction that unitarity imposes which control the low energy $\gamma\gamma\to\pi\pi$ process. These final state interactions are particularly important in the $I=0$ channel. These appear in the reduced amplitudes $\mathcal{R_B}^{I}_{J\lambda}(s)$ above and defined in the Appendix Eq.~(\ref{eq:RLB;fit}).

\section{Pion Polarizabilities}\label{sec:3}
\subsection{Pion Polarizabilities from Dispersion Relations}\label{sec:3;1}
\begin{figure}[th]
\includegraphics[width=0.5\textwidth,height=0.35\textheight]{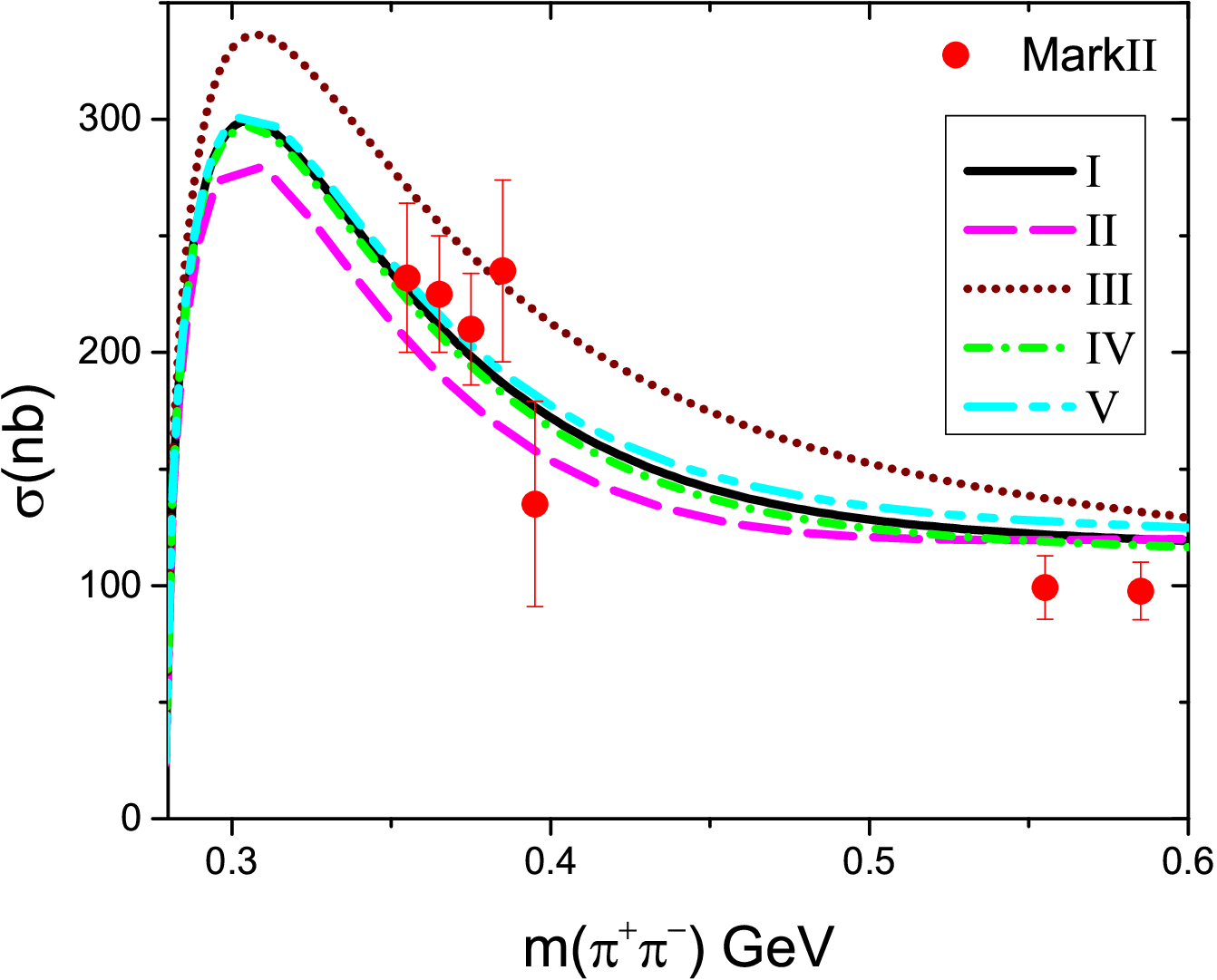}
\includegraphics[width=0.5\textwidth,height=0.35\textheight]{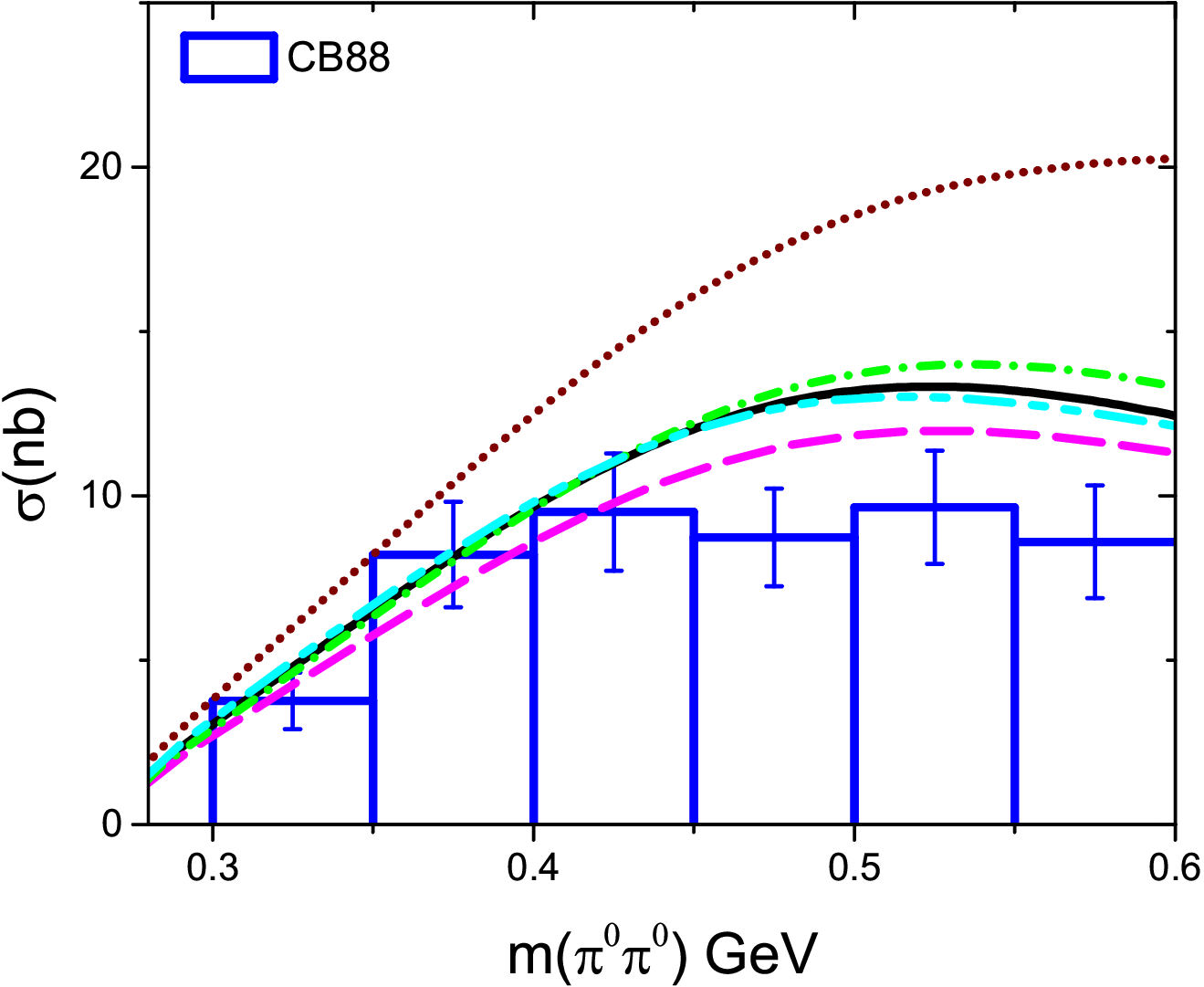}
\caption{\label{Fig:sol;cs} The fits to the $\gamma\gamma\to\pi\pi$  cross-section of the  Models~I-V defined in the text.
The Mark~II~\cite{MarkII} $\pi^+\pi^-$ data are integrated over $|\cos \theta|\,\le\,0.6$, while the Crystal Ball $\pi^0\pi^0$ results~\cite{CB88} are for $|\cos \theta|\,\le\,0.8$. Note the scale of the cross-sections on the left and right differ by more than an order of magnitude.}
\end{figure}

All the Omn$\grave{e}$s functions of Eqs.~(\ref{eq:F;ampS},\ref{eq:F;ampJ}), are fixed from our previous analysis \cite{DLY-MRP14}.
For Left Hand Cut contributions we use the  \lq single particle exchange' model of Sect.~2.2. This should provide an adequate representation at low energies of the effect of even multiparticle exchange, like $2\pi$, $3\pi$, {\it etc.} To get an idea of the range of values  for the polarizabilities we make a series of assumptions, motivated by experimental and theoretical results:
These define Models I-V.
\begin{itemize}
\item Model~I is defined by setting  $(\alpha_1-\beta_1)_{\pi^+}=({4.0\pm1.2\pm1.4})\times10^{-4}$fm$^3$, as given by the latest experiment~\cite{COMPASS2014prl}. We then obtain all the amplitudes and pion polarizability;
\item
Model~II sets $(\alpha_1-\beta_1)_{\pi^+}=0$;
\item Model~III is defined by setting  $(11.6\pm1.5\pm3.0\pm0.5)\times 10^{-4}$fm$^3$ from \cite{Ahrens2005}. This accords with the value of $13.0\times10^{-4}$fm$^3$, as calculated by \cite{Filkov06};

\item Models~IV and V are defined by setting $(\alpha_1-\beta_1)_{\pi^+}=4.0\times10^{-4}$fm$^3$, but fixing the \lq effective' tensor exchange mass ($M_T$) to be 0.8~GeV and 3~GeV, respectively, rather than 1.4~GeV as in Models I-III.
\end{itemize}
The estimates of the polarizability for each of these Models are shown in Table~\ref{tab:p}.
The cross-sections for charged and neutral dipion production from these Models are shown in Fig.~\ref{Fig:sol;cs}.

\begin{table}[t]
{\footnotesize
\begin{center}
\vspace{-0.0cm}
\hspace{-0.0cm}
\begin{tabular}  {|c|c|c|c|c|c|c|c|}
\hline
Polarizabilities                & Model~I        &  Model~II      & Model~III      & Model~IV      & Model~V                  & ChPT +    \rbox \\ [-2mm]
$\lambda=0$ & & & & & &Resonance Model \rbox  \\[0.5mm] \hline
$(\alpha_1-\beta_1)_{\pi^+}$&$\mathbf{4.0\pm1.2\pm1.4}$& $\mathbf{0.0}$     & $\mathbf {11.6 }$        &$\mathbf{ 4.0}$         & $\mathbf{4.0}$                           & 5.7$\pm$1.0       \rbox \\[0.5mm] \hline
$(\alpha_2-\beta_2)_{\pi^+}$& 15.7$\pm$1.1 & 13.0$\pm$1.1 & 20.9$\pm$1.1 & 13.2$\pm$3.4& 18.1$\pm$2.5   & 16.2[21.6]        \rbox \\[0.5mm] \hline
$(\alpha_1-\beta_1)_{\pi^0}$& -0.9$\pm$0.2 & -0.8$\pm$0.1 & -1.1$\pm$0.2 &-0.8$\pm$0.2 &-1.0$\pm$0.2    & -1.9$\pm$0.2      \rbox \\[0.5mm] \hline
$(\alpha_2-\beta_2)_{\pi^0}$& 20.6$\pm$0.8 & 17.8$\pm$0.8 & 26.0$\pm$0.8 &18.6$\pm$2.4 & 22.4$\pm$1.8   & 37.6$\pm$3.3      \rbox \\[0.5mm] \hline
 $\lambda=2$            &              &              &              &             &                &                   \rbox \\[0.5mm] \hline
$(\alpha_1+\beta_1)_{\pi^+}$& 0.26$\pm$0.07& 0.26$\pm$0.07& 0.26$\pm$0.07&0.17$\pm$0.51& 0.42$\pm$0.22  & 0.16[0.16]        \rbox \\[0.5mm] \hline
$(\alpha_2+\beta_2)_{\pi^+}$& -1.4$\pm$0.5 & -1.4$\pm$0.5 & -1.4$\pm$0.5 & -0.9$\pm$3.5& -2.4$\pm$1.5   & -0.001            \rbox \\[0.5mm] \hline
$(\alpha_1+\beta_1)_{\pi^0}$& 0.60$\pm$0.06& 0.60$\pm$0.06& 0.60$\pm$0.06&-0.04$\pm$0.52& 0.90$\pm$0.17 & 1.1$\pm$3.3       \rbox \\[0.5mm] \hline
$(\alpha_2+\beta_2)_{\pi^0}$& -3.7$\pm$0.4 & -3.7$\pm$0.4 & -3.7$\pm$0.4 & 0.4$\pm$3.4 &-5.5$\pm$1.1    & 0.04              \rbox \\[0.5mm] \hline
\end{tabular}
\caption{\label{tab:p}Polarizabilities predicted in Models I-V defined in the text. The highlighted numbers are inputs specifying the particular Model in that column. The final column is for a ChPT+Resonance model. The $\pi^+$ results are from~\cite{Gasser06}, while those for $\pi^0$ are from \cite{Gasser05} and in square brackets from \cite{Bijnens1997}. The units of dipole and quadrupole polarizabilities are in units of  $10^{-4}$fm$^3$ and $10^{-4}$fm$^5$, respectively. $\lambda$ is the total helicity of the two photon system. }
\end{center}
}
%\vspace{-5mm}
\end{table}

 What these results teach are summarized here:
\begin{itemize}
\item{} The first thing to note from Fig.~2 is that the Model~III input of $(\alpha_1-\beta_1)_{\pi^+}= 11.6\times 10^{-4}$fm$^3$ of \cite{Ahrens2005}) is excluded by the $\gamma\gamma\to\pi^0\pi^0$ dataset of Crystal Ball~\cite{CB88}. Thus we do not consider Model~III further.

\item{} Models I, II, IV and  V all essentially predict $(\alpha_1-\beta_1)_{\pi^0}=(-0.9\pm0.2)\times 10^{-4}$fm$^{3}$. This is automatically fixed by constraints of the Adler zero and the input of $(\alpha_1-\beta_1)_{\pi^+}$, see Eq.~(\ref{eq:F;bI0I2},\ref{eq:Po;R}). Otherwise, it is model independent.

\item{} The relation between $(\alpha_1-\beta_1)$ for the $\pi^\pm$ and $\pi^0$ makes it possible to constrain the charged pion polarizability from $\gamma\gamma\to\pi^0\pi^0$ measurements and vice versa. In fact our once or unsubtracted dispersion relations give a strong correlation between  the two photon cross-sections and all helicity zero  polarizabilities, fixing one precisely is sufficient to calculate all the others. The helicity two polarizabilities are fixed, as in Table~1.

\item{} An attempt to reconcile the predictions in the rightmost column of Table~\ref{tab:p} from  Chiral Perturbation Theory to ${\cal O}(p^6)$ with data was carried out by Pasquini, Drechsel and Scherer~\cite{Pasquini08} a decade ago. This gave a very wide range of values for the low energy $\gamma\gamma$ cross-section. This range is explored in  more detail here.

\item{} We find our prediction for $(\alpha_2-\beta_2)_{\pi^0} \simeq 20\times 10^{-4}$fm$^{5}$ is only half that predicted by the ChPT plus Resonance model~\cite{Gasser05}. In contrast, we find $(\alpha_2+\beta_2)_{\pi^+,\pi^0}$ are somewhat larger than other models. The reason is that these are particularly sensitive to LHC contributions from particle exchanges not covered by $\rho$, $\omega$, $b_1$, $h_1$ and $a_1$ --- see how they depend on variations in the mass of the effective tensor exchange between 0.8, 1.4  and 3 GeV (Models~IV, I, V).  Moreover our Omn$\grave{e}$s function differs from other models for the $I=2$ $D$-wave, as we use the phase and they use the phase shift~\cite{Moussallam10}. As discussed earlier \cite{DLY-MRP14}, the phase is quite different from the phase shift for isospin two $D$-waves.

\item{} We obtain $(\alpha_2-\beta_2)_{\pi^+}=15.7\pm1.1\times 10^{-4}$fm$^{5}$ in Model I.
This value is rather close to that in \cite{Phillips11} from their sum rule for the $I=2$ quadrupole polarizabilities deduced using the Roy-Steiner equations. This supports Model I.

\item{}We also note that in Models~ II and III the helicity-two polarizability does not change, as these depend on $D$-waves and $b^I$ is the subtraction constant for the  $S$-wave.

\end{itemize}

\subsection{Error Correlations between Polarizabilities and $\gamma\gamma$ Cross-Sections}\label{sec:3;1}
\begin{figure}[htbp]
\includegraphics[width=0.5\textwidth,height=0.3\textheight]{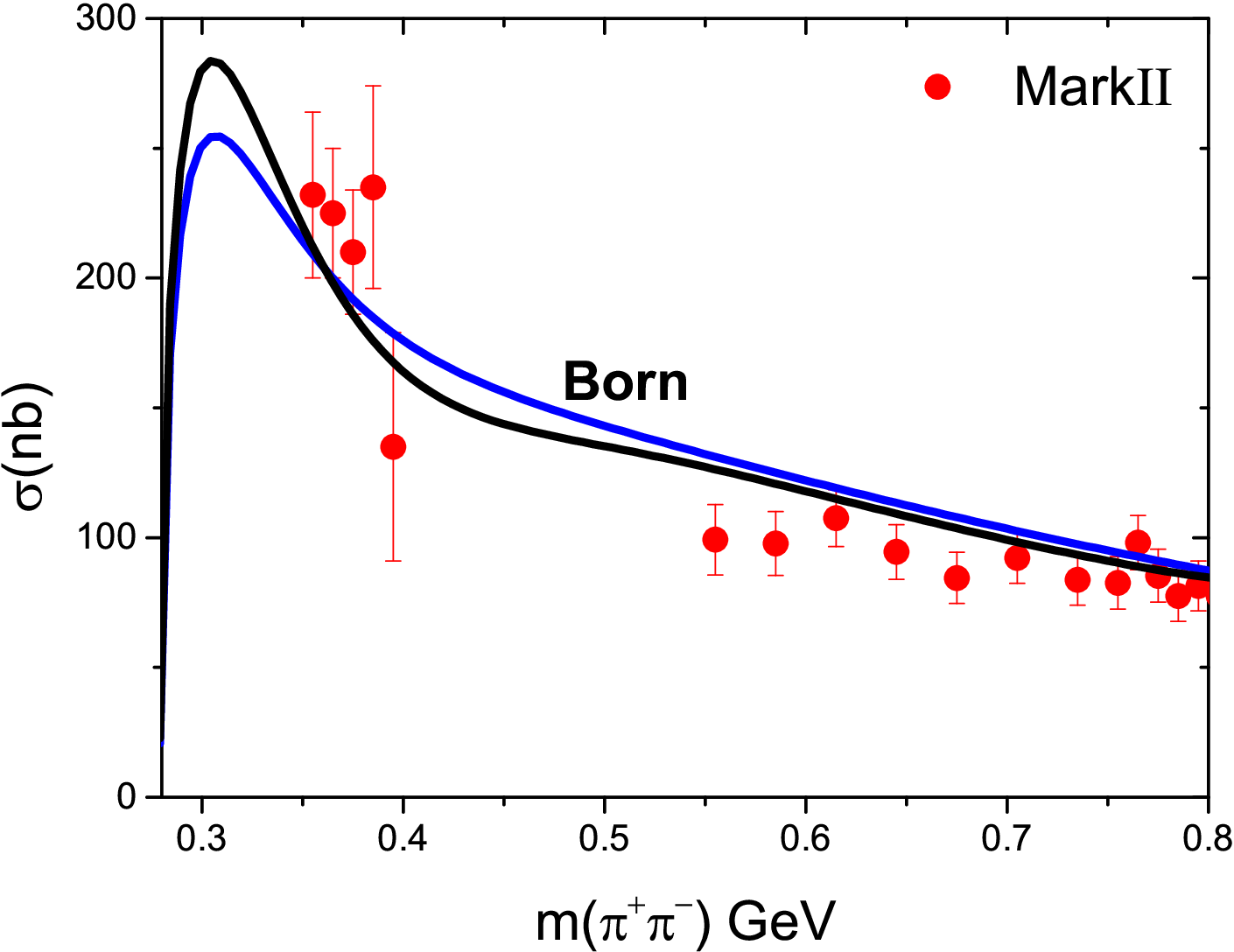}
\includegraphics[width=0.5\textwidth,height=0.3\textheight]{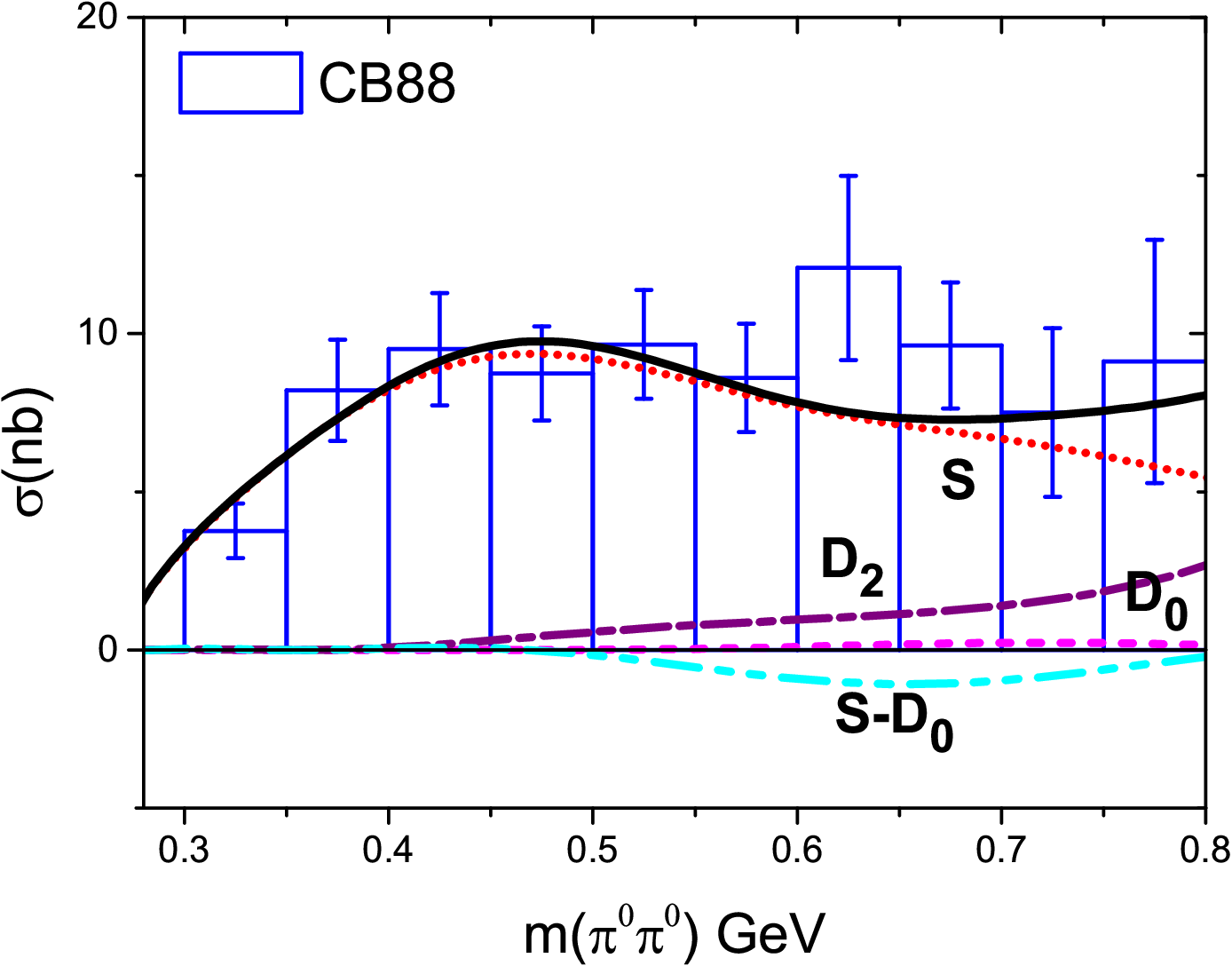}
\caption{\label{Fig:cs} The comparison of the Born terms, full amplitudes of $\gamma\gamma\to\pi^+\pi^-$ and the contribution of each partial wave to $\gamma\gamma\to\pi^0\pi^0$. The data are as shown in Fig.~\ref{Fig:sol;cs}
 The solid black line is the full amplitude from the Amplitude Analysis~\cite{DLY-MRP14}. Note the differing scales of the cross-sections on the left and right.
Since the maximum value of $|\cos\theta| = z = 0.8$ for the neutral pion data, the  $J \le 2$ partial wave contributions not only come from  $|S|^2$ and $|D_\lambda|^2$ (labeled for simplicity without their modulus squared), but also the $S-D_0$ interference, which is negative.  }
\end{figure}

Now let us give an estimate of the uncertainties by investigating  the relation between polarizabilities and the $\gamma\gamma$ cross-sections directly.
The helicity 0 and/or 2 amplitudes of charged and neutral pion production are given as
\bea
F^{c}_{++}(s,\theta,\phi)&=&\left(B_S(s)-s[\sqrt{2/3}b^0\Omega^0_S(s)+\sqrt{1/3}b^2\Omega^2_S(s)] \right.\no\\[1mm]
&&\left.-s^2[\sqrt{2/3}\Omega^0_S(s) R^0_{00}(s)+\sqrt{1/3}\Omega^2_S(s) R^2_{00}(s)]\right)\sqrt{\frac{1}{4\pi}} \no\\[1mm]
&+&\left(B_{D0}(s)-s^2(s-4m_\pi^2)[\sqrt{2/3}\Omega^0_D(s) R^0_{D0}(s)+\sqrt{1/3}\Omega^2_D(s) R^2_{D0}(s)]\right)Y_{D0}(\theta,\phi)\no\\[1mm]
&+&\sum_{J\geq4}B_{J0}(s)Y_{J0}(\theta,\phi)\;,\no\\[3mm]
F^{c}_{+-}(s,\theta,\phi)&=&\left(B_{D2}(s)-s(s-4m_\pi^2)[\sqrt{2/3}\Omega^0_D(s) R^0_{D2}(s)+\sqrt{1/3}\Omega^2_D(s) R^2_{D2}(s)]\right)Y_{D2}(\theta,\phi)\no\\[1mm]
&+&\sum_{J\geq4}B_{J2}(s)Y_{J2}(\theta,\phi)\;,\no\\
%\label{eq:F;ampphyc}
%\eea
%\newpage
%\bea
F^{n}_{++}(s,\theta,\phi)&=&\left(s[-\sqrt{1/3}b^0\Omega^0_S(s)+\sqrt{2/3}b^2\Omega^2_S(s)] \right. \no\\[1mm]
&&\left.+s^2[-\sqrt{1/3}\Omega^0_S(s) R^0_{00}(s)+\sqrt{2/3}\Omega^2_S(s) R^2_{00}(s)]\right)\sqrt{\frac{1}{4\pi}} \no\\[1mm]
&+&\left(s^2(s-4m_\pi^2)[-\sqrt{1/3}\Omega^0_D(s) R^0_{D0}(s)+\sqrt{2/3}\Omega^2_D(s) R^2_{D0}(s)]\right)Y_{D0}(\theta,\phi)\no\\[3mm]
F^{n}_{+-}(s,\theta,\phi)&=&\left(s(s-4m_\pi^2)[-\sqrt{1/3}\Omega^0_D(s) R^0_{D2}(s)+\sqrt{2/3}\Omega^2_D(s) R^2_{D2}(s)]\right)Y_{D2}(\theta,\phi)\;.\label{eq:F;ampphy}
\eea
For $\gamma\gamma\to\pi^+\pi^-$, because of the threshold factors, the LHCs will contribute just a little to the charged pion polarizability compared to the effect of final state interaction that modifiy the Born terms (mainly  S, $D_2$ waves) in the low energy region. For $\gamma\gamma\to\pi^0\pi^0$, the $S$-wave dominates at low energy and the contribution of higher partial waves is small. The details are shown in Fig.\ref{Fig:cs}.

As seen in Eq.~(\ref{eq:Po;R}), it is $R^0_{00}$ and $R^2_{00}$ are the dominant part of the polarizabilities $(\alpha_2-\beta_2)_{\pi^+,\pi^0}$ and $R^0_{22}$ and $R^2_{22}$ dominate for $(\alpha_2+\beta_2)_{\pi^+,\pi^0}$. That is to say, we can ignore the derivative part of the Omn$\grave{e}$s functions.
Keeping these in mind and noting that when $s$ is small the value of Omn$\grave{e}$s functions, as defined in Eq.~(\ref{eq:Omnes}), are very close to one, these can be set to unity in Eqs.~(\ref{eq:F;ampphy}) to make the error estimate. Of course, we use the full Omn$\grave{e}$s functions in the $R^I_{J\lambda}$ functions in making the predictions in Table~1.

Unfortunately, the measurement of the two photon production of mesons do not cover the full angular range. This is limited to $|\cos\theta|\le z$. In $e^+e^-$ colliders, $z$ is typically 0.6-0.7 for charged pions and 0.8 for $\pi^0\pi^0$. The GlueX experiment will produce good angular coverage for $40^o < \theta < 140^o$ according to ~\cite{Jlabproposal}, so $z=0.77$. Consequently, the differential cross-sections are integrated up to $\cos\theta = z$ to give $\sigma^{c,n}(s,z)$ with uncertainties $\Delta\sigma^{nc,n}(s,z)$. We can readily estimate the relative errors between polarizability and cross-sections from Eq.~(\ref{eq:Po;R}) to be:

\bea
\Big{|}\frac{\Delta\sigma^{c}(s,z)}{\sigma^{c}(s,z)}\Big{|}&\doteq&\Big{|}\frac{\Delta(\alpha_1-\beta_1)_{\pi^+}}{(\alpha_1-\beta_1)_{\pi^+}}\Big{|} C_{(\alpha_1-\beta_1)_{\pi^+}}(s,z)
+\Big{|}\frac{\Delta(\alpha_2-\beta_2)_{\pi^+}}{(\alpha_2-\beta_2)_{\pi^+}}\Big{|} C_{(\alpha_2-\beta_2)_{\pi^+}}(s,z)\no\\[2mm]
&+&\Big{|}\frac{\Delta(\alpha_1+\beta_1)_{\pi^+}}{(\alpha_1+\beta_1)_{\pi^+}}\Big{|} C_{(\alpha_1+\beta_1)_{\pi^+}}(s,z)
+\Big{|}\frac{\Delta(\alpha_2+\beta_2)_{\pi^+}}{(\alpha_2+\beta_2)_{\pi^+}}\Big{|} C_{(\alpha_2+\beta_2)_{\pi^+}}(s,z)\;,\no\\[15mm]
\Big{|}\frac{\Delta\sigma^{n}(s,z)}{\sigma^{n}(s,z)}\Big{|}&\doteq&\Big{|}\frac{\Delta(\alpha_1-\beta_1)_{\pi^0}}{(\alpha_1-\beta_1)_{\pi^0}}\Big{|} C_{(\alpha_1-\beta_1)_{\pi^0}}(s,z)
+\Big{|}\frac{\Delta(\alpha_2-\beta_2)_{\pi^0}}{(\alpha_2-\beta_2)_{\pi^0}}\Big{|} C_{(\alpha_2-\beta_2)_{\pi^0}}(s,z)\no\\[2mm]
&+&\Big{|}\frac{\Delta(\alpha_1+\beta_1)_{\pi^0}}{(\alpha_1+\beta_1)_{\pi^0}}\Big{|} C_{(\alpha_1+\beta_1)_{\pi^0}}(s,z)
+\Big{|}\frac{\Delta(\alpha_2+\beta_2)_{\pi^0}}{(\alpha_2+\beta_2)_{\pi^0}}\Big{|} C_{(\alpha_2+\beta_2)_{\pi^0}}(s,z)\;.\label{eq:P;C}
\eea
where the $C$-functions are given by
\bea
C_{(\alpha_1-\beta_1)_{\pi^+}}(s,z)&=&\Big{|}\frac{2\pi\alpha m_\pi \rho(s)B_S(s) z(\alpha_1-\beta_1)_{\pi^+}}{ \sigma_B^c(s,z) }\Big{|}\;,\no\\[2.5mm]
 C_{(\alpha_2-\beta_2)_{\pi^+}}(s,z)&=&\Big{|}\frac{s\pi\alpha m_\pi \rho(s) B_S(s) z(\alpha_2-\beta_2)_{\pi^+}}{6 \sigma_B^c(s,z) }\Big{|}\;,\no\\[2.5mm]
C_{(\alpha_1+\beta_1)_{\pi^+}}(s,z)&=&\Big{|}\frac{\pi\alpha(s-4m_\pi^2)\rho(s) B_{D2}(s) z(15 - 10 z^2 + 3 z^4)(\alpha_1+\beta_1)_{\pi^+}}{4\sqrt{30} m_\pi \sigma_B^c(s,z) }\Big{|}\;,\no\\[2.5mm]
C_{(\alpha_2+\beta_2)_{\pi^+}}(s,z)&=&\Big{|}\frac{\pi\alpha(s-4m_\pi^2) m_\pi \rho(s) B_{D2}(s) z(15 - 10 z^2 + 3 z^4)(\alpha_2+\beta_2)_{\pi^+}}{12\sqrt{30}  \sigma_B^c(s,z) }\Big{|}\;,\no\\[2.5mm]
C_{(\alpha_1-\beta_1)_{\pi^0}}(s,z)&=&
\Big{|}\frac{s m_\pi (\alpha_1-\beta_1)_{\pi^0}}{2\alpha F^{n}_{S}(s) }\Big{|}\;,\no\\[2.5mm]
C_{(\alpha_2-\beta_2)_{\pi^0}}(s,z)&=&\Big{|}\frac{s^2 m_\pi (\alpha_2-\beta_2)_{\pi^0}}{ 24\alpha F^{n}_{S}(s) }\Big{|}\;,\no\\[2.5mm]
C_{(\alpha_1+\beta_1)_{\pi^0}}(s,z)&=&\Big{|}\frac{s(s-4m_\pi^2)(\alpha_1+\beta_1)_{\pi^0}F^{n}_{D2}(s)(15 - 10 Z^2 + 3 Z^4)}{ 16\sqrt{30}\alpha m_\pi F^{n}_{S}(s)^2 }\Big{|}\;, \no\\[2.5mm]
C_{(\alpha_2+\beta_2)_{\pi^0}}(s,z)&=&
\Big{|}\frac{s(s-4m_\pi^2)m_\pi (\alpha_2+\beta_2)_{\pi^0} F^{n}_{D2}(s)(15 - 10 z^2 + 3 z^4)}{ 48\sqrt{30}\alpha F^{n}_{S}(s)^2  }\Big{|}^2\;.\label{eq:C}
\eea
The Eqs.(\ref{eq:C}) involve the integrated Born cross-section, $\sigma_B(s,z)$, which with  $\rho=\rho(s)$ of Eq.~(\ref{eq:X;rho}), is given by
\bea
\sigma_B^c(s,z)&=&\frac{2\pi\alpha^2\rho}{s}\left[ z + \frac{(1-\rho^2)^2 \,z}{1-\rho^2 z^2}-\frac{(1-\rho^4)}{2\rho}\ln\left(\frac{1+\rho z}{1-\rho z}\right)\right] \, .
\eea

\begin{figure}[p]
\begin{center}
\includegraphics[width=0.85\textwidth,height=0.85\textheight]{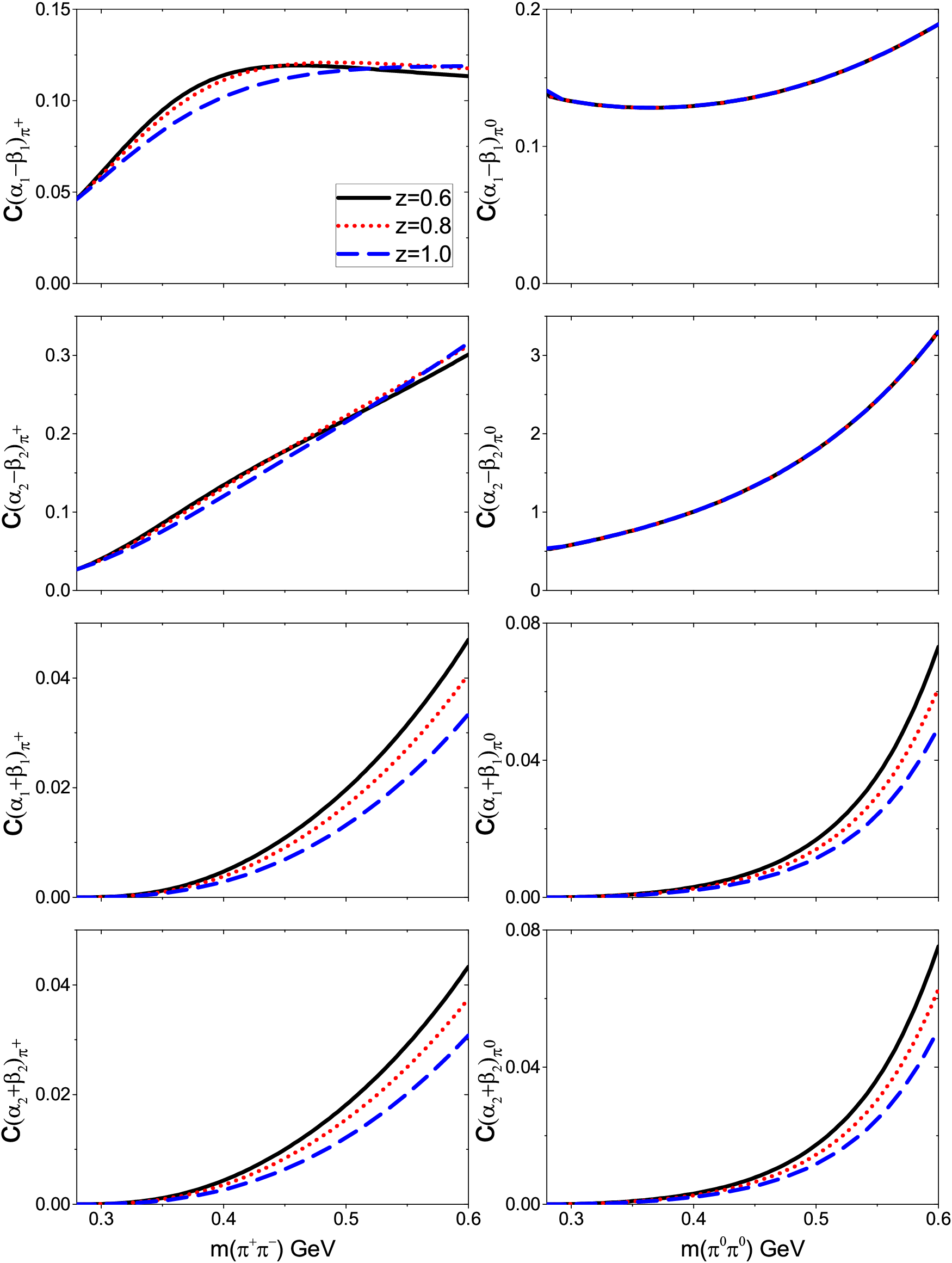}
\caption{\label{Fig:c}The relation between the relative errors of cross-section and polarizability, the $C$-functions are defined in Eq.~(\ref{eq:P;C}). The solid lines are for $\gamma\gamma$ cross-section measured up to  $|\cos\theta|=z$ with $z=0.6$, the dotted lines with $z=0.8$ and the dashed lines with $z=1$.}
\end{center}
\end{figure}

\begin{table}[th]
{\footnotesize
\begin{center}
\vspace{-0.0cm}
\hspace{-0.0cm}
\begin{tabular}  {|c|c|c|c|}
\hline
Polarizability                & For an uncertainty of        &  Accuracy required of $\gamma\gamma\to\pi\pi$  & Uncertainty required in the \rbox \\ [-2mm]
 & & cross-section at 450~MeV & integrated cross-section \rbox  \\[0.5mm] \hline
$(\alpha_1-\beta_1)_{\pi^+}$& 100\%& 10\% & 20~nb\rbox \\[0.5mm] \hline
$(\alpha_2-\beta_2)_{\pi^+}$& 100\%& 17\% & 34~nb \rbox \\[0.5mm] \hline
$(\alpha_1-\beta_1)_{\pi^0}$& 100\%& 13\% & 1.2 ~nb\rbox \\[0.5mm] \hline
$(\alpha_2-\beta_2)_{\pi^0}$& 100\%& 132\%& 12~nb \rbox \\[0.5mm] \hline
$(\alpha_1+\beta_1)_{\pi^+}$& 100\%& 1\% & 2~nb \rbox \\[0.5mm] \hline
$(\alpha_2+\beta_2)_{\pi^+}$& 100\%& 1\% & 2~nb\rbox \\[0.5mm] \hline
$(\alpha_1+\beta_1)_{\pi^0}$& 100\%& 1\% & 0.08~nb\rbox \\[0.5mm] \hline
$(\alpha_2+\beta_2)_{\pi^0}$& 100\%& 1\% & 0.07~nb \rbox \\[0.5mm] \hline
\end{tabular}
\caption{\label{tab:accuracy} To determine each polarizability listed with an uncertainty of 100\%, the corresponding (charged or neutral pion) cross-section for $\gamma\gamma\to\pi\pi$ has to be measured at 450~MeV (as an example) to the accuracy tabulated for $z=0.77$ for charged and neutral pions, where GlueX is expected to have good angular coverage~\cite{Jlabproposal}. At other energies the percentage accuracies can be read off from the graphs in Fig.~4. }
\end{center}
}
\end{table}

A general estimate of the error correlations for each polarizability in Table~1 is shown in Fig~\ref{Fig:c}. We see that if we want to fix the uncertainty of the polarizability at 100 percent, the accuracy of the $\gamma\gamma\to\pi\pi$ cross-section at $\sqrt{s}$ of 450~MeV (when $z=0.6$) for charged pions, and with $z=0.8$ for neutral pions to the precision listed in Table~2. The values at other energies can be read off the plots in Fig.~4.
Among these only the value of the $C_{(\alpha_2-\beta_2)_{\pi^0}}$ is large, we therefore suggest that experiment measures the $\gamma\gamma\to\pi^0\pi^0$ cross-section to fix $(\alpha_2-\beta_2)_{\pi^0}$. The values of $C$-function of helicity-two polarizabilities,  $(\alpha_1+\beta_1)_{\pi^+,\pi^0}$ and $(\alpha_2+\beta_2)_{\pi^+,\pi^0}$, have larger values for the neutral pion. Neverthless they are especially small. The reason is that they are related to $D$-waves and in the low energy region $D$-waves are strongly suppressed by the threshold factors $s^n(s-4m_\pi^2)^{J/2}$,  thus they hardly contribute to the cross-section.
%COMMENTED OUT: {\it This is also compatible with what we discussed in the last section: helicity 2 polarizabilities are not sensitive to the LHCs. Together with the discussion before, fixing helicity-zero (but not helicity-two) polarizability would be helpful to determine the LHC contribution.
%We note that the value of $C$-function of $\gamma\gamma\to\pi^0\pi^0$ is larger than that of $\gamma\gamma\to\pi^+\pi^-$, except for that of $(\alpha_1+\beta_1)_{\pi^0}$, $(\alpha_2+\beta_2)_{\pi^0}$. But again when energy increases, the $C$-function value of $\gamma\gamma\to\pi^0\pi^0$ will be larger.
%This is because Born term dominates the low energy cross-section for $\gamma\gamma\to\pi^+\pi^-$, and it is only the difference between the full amplitude and the Born term that is related to the polarizability.}
We also find that the $C$-functions increase as the energy goes higher, this is an important observation as it shows an Amplitude Analysis at a little higher energy, away from  threshold, is necessary to determine the polarizabilities. We would suggest that experiments measure the $\gamma\gamma$ cross-sections in the energy range of  $\sqrt{s}\sim 350$ and 600~MeV. Too low the cross-section is not sensitive to the polarizability. Too high then our analysis using Eq.~(\ref{eq:P;C}) is no longer valid, as the Omn$\grave{e}$s functions change much more, making the correlation between polarizability and cross-section uncertainties more complicated.

\section{Conclusion}
In this paper we give our estimate of pion polarizabilities based on our earlier Amplitude Analysis~\cite{DLY-MRP14}. Our use of once subtracted dispersion relations for the $S$-waves and unsubtracted for all other waves provides a tighter constraint between the two photon cross-sections in the low energy region. This correlates the charged and neutral pion cross-sections and the helicity-zero charged and neutral pion polarizabilities. Confirming any of these quantities with precision fixes the others. The polarizabilities for a number of differing inputs are listed in Table \ref{tab:p} as Models I-V. The correlation of relative errors between pion polarizability and two photon cross-section are shown in Fig.~\ref{Fig:c} and summarized in Table~2 at $\sqrt{s}$ of 450~MeV. Model I is the most likely based on the latest measured value of $(\alpha_1-\beta_1)_{\pi^+}$ from COMPASS~\cite{COMPASS2014prl}. The helicity-zero polarizabilities are rather stable as known final state interactions modifying the Born terms make the dominant contribution. They are the least sensitive to Chiral/Resonance models. Consequently, one of the first $\gamma\gamma$ measurements should be for charged pion production to confirm the COMPASS value for $(\alpha_1-\beta_1)_{\pi^+}$. This should take advantage, for instance, of the good angular coverage of GlueX~\cite{Jlabproposal}. Then the $\pi^+\pi^-$ cross-section must be measured to better than $\pm 2.2$~nb  to fix this polarizability to an accuracy of 10\%. With this value known, then  $(\alpha_1-\beta_1)_{\pi^0}=(-0.9\pm0.2)\times 10^{-4}$fm$^{3}$ is fixed in a model independent way. Only experimental input on $(\alpha_1-\beta_1)_{\pi^+}$ and the position of the Adler zero will constrain it.
Indeed, we find that the helicity-zero polarizability is much more sensitive to the $\gamma\gamma$ cross-section than those of helicity-two, making them easier to measure in experiment and easier to connect using dispersion relations.

%\newpage
The largest uncertainties come from ill-determined left hand cut contributions to the dispersion relations for the $\gamma\gamma$ partial waves. These are reflected in the what we call the $C$-functions, Eq.~(\ref{eq:C}), that enter in the correlation between polarizabilities and two photon cross-sections. These are very small around threshold, but increase when the energy goes higher. As a consequence we stress that the best region to measure the $\gamma\gamma$ cross-sections is at the {\it intermediate} energy region of $\sqrt{s}$ from 350 to 600~MeV.
Of the helicity-two quantities we find that $(\alpha_2-\beta_2)_{\pi^0}$ is the easiest polarizability to  fix by measuring the  $\gamma\gamma\to\pi^0\pi^0$ cross-section. What is more, it is the least sensitive to variations of the left hand cut, thus easier for theory to check.
Future experiments at COMPASS at CERN, and GlueX at Jefferson Lab are the most suitable for studying pion polarizabilities.

\vspace{1cm}

\section*{Acknowledgments}
\noindent
We thank U.-G. Mei{\ss}ner for reading the paper and for his suggestions.
This work is supported in part by the DFG (SFB/TR 110, ``Symmetries and the Emergence of Structure in QCD''). We acknowledge support from Indiana University College of Arts and Sciences, and from the U.S. Department of Energy, Office of Science, Office of Nuclear Physics under contract DE-AC05-06OR23177 that funds Jefferson Lab research.

\vspace{1cm}

%\newpage
\appendix
\setcounter{equation}{0}
\setcounter{table}{0}
\renewcommand{\theequation}{\Alph{section}.\arabic{equation}}
\renewcommand{\thetable}{\Alph{section}.\arabic{table}}

\section{Definition of Reduced Amplitudes}\label{app:A}
\noindent
It is convenient to determine such functions:
\bea
\mathcal{R_L}^{I}_{00}(s) &= & \frac{1}{\pi}\int_L ds'\frac{{\rm Im}\left[ \mathcal{L}^{I}_{00}(s')\right]\Omega^{I}_{0}(s')^{-1} }{s'^2(s'-s)}\;,\no\\
\mathcal{R_L}^{I}_{J\lambda}(s) &= & \frac{1}{\pi}\int_L ds'\frac{ {\rm Im}\left[ \mathcal{L}^{I}_{J\lambda}(s')\right]\Omega^{I}_{0}(s')^{-1} }{s'^n(s'-4m_\pi^2)^{J/2}(s'-s)}\;,\;\;\; \text{for}\; J\geq2\;\,,\no\\
\mathcal{R'_L}^{I}_{00}(s) &= & \frac{1}{\pi}\int_L ds'\frac{{\rm Im}\left[ \mathcal{L}^{I}_{00}(s')\right]\Omega^{I}_{0}(s')^{-1} }{s'^2(s'-s)^2}\,,\no\\
\mathcal{R'_L}^{I}_{J\lambda}(s) &= & \frac{1}{\pi}\int_L ds'\frac{ {\rm Im}\left[ \mathcal{L}^{I}_{J\lambda}(s')\right]\Omega^{I}_{0}(s')^{-1} }{s'^n(s'-4m_\pi^2)^{J/2}(s'-s)^2}\;,\label{eq:RL;fit}\\[3mm]
\mathcal{R_B}^{I}_{00}(s) &\equiv& -\frac{1}{\pi}\,\int_R ds'\frac{B^{I}_{S}(s')\,  {\rm Im}\left[ \Omega^{I}_{0}(s')^{-1}\right] }{s'^2(s'-4m_\pi^2)^{J/2}(s'-s)}\;,\no\\
\mathcal{R_B}^{I}_{J\lambda}(s) &\equiv& -\frac{1}{\pi}\,\int_R ds'\frac{B^{I}_{J\lambda}(s')\,  {\rm Im}\left[ \Omega^{I}_{J}(s')^{-1}\right] }{s'^n(s'-4m_\pi^2)^{J/2}(s'-s)}\;,\;\;\; \text{for}\; J\geq2\;\,,\no\\
\mathcal{R'_B}^{I}_{00}(s) &\equiv& -\frac{1}{\pi}\,\int_R ds'\frac{B^{I}_{S}(s')\,  {\rm Im}\left[ \Omega^{I}_{0}(s')^{-1}\right] }{s'^2(s'-4m_\pi^2)^{J/2}(s'-s)^2}\;,\no\\
\mathcal{R'_B}^{I}_{J\lambda}(s) &\equiv& -\frac{1}{\pi}\,\int_R ds'\frac{B^{I}_{J\lambda}(s')\,  {\rm Im}\left[ \Omega^{I}_{J}(s')^{-1}\right] }{s'^n(s'-4m_\pi^2)^{J/2}(s'-s)^2}\;.\label{eq:RLB;fit}
\eea
and
\bea
\mathcal{R}^{I}_{J\lambda}(s)&=&\mathcal{R_L}^{I}_{J\lambda}(s)+\mathcal{R_B}^{I}_{J\lambda}(s) \nonumber\\
\mathcal{R'}^{I}_{J\lambda}(s)&=&\mathcal{R'_L}^{I}_{J\lambda}(s)+\mathcal{R'_B}^{I}_{J\lambda}(s)
\eea
Note that we have divided out the threshold behaviour factors \lq\lq $s^2$, $s^n(s-4m_\pi^2)^{J/2}$ '' in $\mathcal{R}^{I}_{J\lambda}(s)$. These $\mathcal{R'}^{I}_{J\lambda}(s)$ functions describe the amplitudes well near threshold.
As an estimate we use single resonance exchange, shown in Eq.~(\ref{eq:L;RChT}), to simulate the left hand cuts and calculate the amplitudes at low energy region.

%%%%%%%%%%%%%%%%%%%%%%%%%%%%%%%%%%%%%%%%%%%%%%%%%%%%%%%%%%%%%%%%%%%%%%%%%%%%%%%%%%%%%%%%%%%%%%%%%%%
%\newpage

%%%%%%%%%%%%%%%%%%%%%%%%%%%%%%%%%%%%%%%%%%%%%%%%%%%%%%%%%%%%%%%%%%%%%%%%%%%%%%%%%%%%%%%%%%%%%%%%%%%

\end{document}